\documentclass[journal]{vgtc}                     

\onlineid{1594}

\vgtccategory{Research}

\vgtcpapertype{Representations \& Interaction}

\title{``I Came Across a Junk'': Understanding Design Flaws of Data Visualization from the Public's Perspective}

\author{Xingyu Lan, Yu Liu}

\authorfooter{
\item
Xingyu Lan is with Fudan University and a member of the Research Group of Computational and AI Communication at Institute for Global Communications and Integrated Media; 
Yu Liu is a MSc student in University of Edinburgh\\
E-mail: \{xingyulan96,coraline.liu.dataviz\}@gmail.com. Xingyu Lan is the corresponding author.
}

\abstract{%
  The visualization community has a rich history of reflecting upon visualization design flaws. Although research in this area has remained lively, we believe it is essential to continuously revisit this classic and critical topic in visualization research by incorporating more empirical evidence from diverse sources, characterizing new design flaws, building more systematic theoretical frameworks, and understanding the underlying reasons for these flaws.
  To address the above gaps, this work investigated visualization design flaws through the lens of the public, constructed a framework to summarize and categorize the identified flaws, and explored why these flaws occur. Specifically, we analyzed 2227 flawed data visualizations collected from an online gallery and derived a design task-associated taxonomy containing 76 specific design flaws. These flaws were further classified into three high-level categories (i.e., misinformation, uninformativeness, unsociability) and ten subcategories (e.g., inaccuracy, unfairness, ambiguity). Next, we organized five focus groups to explore why these design flaws occur and identified seven causes of the flaws. Finally, we proposed a research agenda for combating visualization design flaws and summarize nine research opportunities.
}


\graphicspath{{figs/}{figures/}{pictures/}{images/}{./}} 

\usepackage{tabu}                      
\usepackage{booktabs}                  
\usepackage{lipsum}                    
\usepackage{mwe}                       

\newcommand{\etal}{et~al.~} 
\newcommand{\ie}{i.e.,~}
\newcommand{\eg}{e.g.,~}

\usepackage{microtype}                 
\PassOptionsToPackage{warn}{textcomp}  
\usepackage{textcomp}                  
\usepackage{mathptmx}                  
\usepackage{times}                     
\usepackage{cite}                      
\usepackage{tabu}                      
\usepackage{booktabs}                  
\usepackage{enumitem}
\usepackage{url}
\usepackage{soul}
\makeatletter
\g@addto@macro{\UrlBreaks}{\UrlOrds}
\makeatother
\usepackage{graphicx}
\usepackage{wrapfig}
\usepackage[T1]{fontenc}
\usepackage{amsmath}  
\usepackage{multirow}
\usepackage{tabularx}

\usepackage[svgnames,x11names,table]{xcolor}
\usepackage{soul}
\usepackage{adjustbox}

\definecolor{c1}{HTML}{f9856e}
\definecolor{c2}{HTML}{79acf7}
\definecolor{c3}{HTML}{9ad6b4}

\definecolor{t1}{HTML}{798fda}
\definecolor{t2}{HTML}{9ad6b4}
\definecolor{t3}{HTML}{d87191}
\definecolor{t4}{HTML}{dc806e}
\definecolor{t5}{HTML}{f6d479}

\usepackage{mathptmx}                  

\begin{document}

\firstsection{Introduction}
\maketitle


The visualization community has a long history of reflecting on flawed visualization design. For example, \textit{How to Lie with Statistics}, written by Darrell Huff in 1954, discussed a series of misleading charts collected from newspapers. The book quickly gained global popularity, with over one and a half million copies sold to date.
Later, the proposal of \textit{chart junk} by Edward Tufte~\cite{tufte2001visual} in the 1980s constitutes another milestone. In his book, Tufte criticized overly ornate chart designs and proposed indicators such as the lie factor to help quantify the conciseness of a chart.
These pioneering works and their significant social impact have clearly demonstrated that visualizations have the power to sway perceptions and decisions, and inappropriate visualization can undermine the credibility of data or even erode the public's trust in the field of visualization itself.
Today, such challenges are even more significant: with the democratization of data, everyone is able to create visualizations, and social media makes the dissemination of misinformation unprecedentedly easy.
It is imperative for us, as visualization researchers, to continuously examine flawed visualization designs and to explore this issue more thoroughly from diverse perspectives.

In recent years, research on flawed visualization design has remained lively. For example, Pandey~\etal~\cite{pandey2015deceptive} conducted experiments to investigate the effects of four common distortion techniques specifically (\eg truncated axis, inverted axis). 
By reviewing previous literature, Mcnutt~\etal~\cite{mcnutt2020surfacing} proposed a framework of \textit{visualization mirages} (silent but significant failures in a visualization). Lo~\etal~\cite{lo2022misinformed} coded 1143 misinformative visualizations collected from the web and proposed a taxonomy containing 74 design faults.
However, the discussion on this research direction is far from settled. We believe more studies are still needed to repetitively investigate this critical theme in visualization research. In particular, incorporating more empirical evidence from diverse sources would help reflect on design flaws that have not yet established consensus in our community as well as characterize new design flaws beyond known ones.
For example, Wainer~\cite{wainer1984display} once listed ``showing as few data as possible'' as one rule of bad data display. As a statistician, Wainer believed that increasing the density of information on a graph is more desirable. However, research conducted with the general public found that visualizations with dense information can be overwhelming and annoying~\cite{lan2021smile}.
Correll~\cite{correll2021towards} argued that previous research focused mainly on the flaws of illegibility and data distortion, but ``these two categories are such a small part of what makes a visualization work''. He then coined the term \textit{bullshit visualization} to describe ``charts that do not have even the common decency to intentionally lie but are totally unconcerned about the state of the world or any practical utility.''
This indicates that a more systematic framework that covers a wider variety of problems than just intentional deception is desired.
Last but not the least, in contrast to researchers' strong interest in identifying \textit{what} visualization design flaws are, there has been relatively little investigation into \textit{why} they occur, which may hinder our understanding of how to effectively combat such flaws. 


To address the above gaps, this work aims to investigate visualization design flaws through the lens of the public, construct a framework to summarize and categorize the identified flaws, and explore why these flaws occur. 
Our research is based on an online gallery called WTF Visualizations~\cite{wtf}, which was established in 2013 and has been showcasing flawed visualizations contributed by the public for more than ten years. 
First, we scraped all the visualizations from the gallery, resulting in a corpus containing 2297 images. Then, we cleaned the corpus, coded 2227 visualizations, and analyzed their design flaws by referring to the original comments submitted by their up-loaders. This led to a taxonomy containing 76 identified flaws classified into three high-level categories (\ie uninformativeness, misinformation, unsociability) and ten subcategories (\eg inaccuracy, unfairness, ambiguity). Next, we organized multiple rounds of focus groups to spark discussions around the flawed designs in our corpus, aiming to further infer why the design flaws occur.
Based on the above analyses and findings, we discuss the implications arising from this work and provide suggestions for combating design flaws in data visualization.

To conclude, the contributions of this work include: (i) We analyzed a corpus sourced from the general public, providing additional real-world examples of visualization design flaws across various domains and topics and covering a wide range of issues beyond deliberate deception.
(ii) We proposed a taxonomy that differs from previous research for categorizing visualization design flaws, accompanied by the reporting of newly identified flaws and patterns. (iii) We explored the underlying causes of the identified design flaws and proposed a research agenda for combating them.
All the codes and raw images can be found in our supplemental materials as well as our website. (\url{https://flawviz.github.io/}).
\section{Related Work}

We review prior literature about design flaws in data visualization, visualization design guidelines, and visualization for the general public.

\subsection{Design Flaws of Data Visualization}

Although data visualization is known for its effectiveness in transforming abstract data into interpretable visuals, poorly designed visualizations can have negative side effects~\cite{o2018testing,lauer2020people}. This fact was realized even before visualization became a formal discipline. For example, as early as 1939, engineer and InfoVis pioneer Brinton~\cite{brinton1939graphic} highlighted several design flaws that could distort data in his book \textit{Graphic Presentation}, such as omitting zero values and using disproportionate grids. In the 1950s, journalist Huff~\cite{huff2023lie} published his best-selling book \textit{How to Lie with Statistics}, discussing misleading charts found in newspapers.
In 1983, Tufte~\cite{tufte2001visual} introduced the concept of \textit{chart junk} to criticize visualizations that include unnecessary embellishments. This sparked a debate on chart junk, leading to a series of studies about embellished visualization~\cite{bateman2010useful,borgo2012empirical,andry2021interpreting,haroz2015isotype,alebri2023embellishments}. Additionally, other design methods such as truncated/inverted axis~\cite{szafir2018good,lauer2020people,fan2022annotating,wainer1984display,pandey2015deceptive}, 3D effects~\cite{szafir2018good,cairo2019charts,lauer2020people,ware2019information}, rainbow palettes~\cite{borland2007rainbow,szafir2018good,ware2019information}, and Chernoff faces~\cite{lee2003empirical,morris2000experimental} have also been frequently studied as flawed design choices.
%

However, there is still much room for research and discussion surrounding flawed visualization design. Although researchers have summarized a set of design techniques that are considered inappropriate (\eg ~\cite{tufte2001visual,wainer1984display}), many of them are still in controversy. For instance, although embellishment is considered redundant and inefficient in the context of minimalism, it has been found to be "useful junk" that can make visualizations more memorable~\cite{bateman2010useful}. Similarly, while manipulating the aspect ratio of visualizations has been identified as a common deceptive tactic in the work by Pandey~\etal~\cite{pandey2014persuasive}, Cairo~\cite{cairo2019charts}, a practitioner with extensive experience in creating visualizations, stated that this tactic is not necessarily wrong in certain contexts.
Furthermore, there is a wide range of interpretations and even disagreements regarding the classic concept of chart junk~\cite{parsons2020data,akbaba2021manifesto}.
Secondly, the current discussion about visualization design flaws has been focusing mainly on the distortion of perception, however, as argued by Correll~\cite{correll2021towards,correll2017black} and Lisnic~\etal~\cite{lisnic2023misleading}, perceptual problems are only one category in the larger space of bad visualization. 
Previous work has also found that users evaluate visualization design from multiple angles, such as usability, expressiveness, and aesthetics~\cite{lan2021smile,cawthon2007effect,bresciani2015pitfalls,kauer2021public}, indicating that "bad design" may occur at various levels. 

Given the above gaps, we believe that more incremental work needs to be done to provide additional evidence for what constitutes a flawed visualization from different perspectives and based on different empirical samples. Thus, this work revisits this classic research question by analyzing a gallery contributed by the general public to explore the landscape of visualization design flaws in the eyes of the public.

\subsection{The Formulation of Design Guidelines}
\label{sec:formulation}
To help identify and avoid visualization design flaws, researchers have established a set of rule-based guidelines. For example, in the 1980s, psychologists such as Cleveland and
McGill~\cite{cleveland1984graphical} conducted a series of experiments to compare the performance of different visual channels on low-level perceptual tasks. 
Such rules have helped the intelligent recommendation or generation of visualizations based on input variables~\cite{mackinlay1986automating,moritz2018formalizing}, as well as the linting of errors in visualizations~\cite{chen2021vizlinter,hopkins2020visualint,mcnutt2020surfacing,fan2022annotating,ritchie2019lie}. 
However, currently, although the visualization community has reached a high consensus on some basic design guidelines (\eg the channel of shape is more suitable for encoding categorical variables rather than continuous variables), our knowledge about visualization design is still limited in many aspects.
For example, as reflected by Chen~\etal~\cite{chen2021vizlinter}, their linting system is only able to detect "basic construction errors of visualization". Therefore, they called for future work to extend the coverage of rule categories (especially soft rules concerning expressiveness, aesthetic, or stylistic issues) and to conduct studies to better understand users' judgments of various errors.

Nevertheless, so far, only a few papers have attempted to formulate a comprehensive taxonomy for visualization design flaws.
As one example, Bresciani and Eppler~\cite{bresciani2015pitfalls} categorized common errors in data visualization into three types (\ie cognitive, emotional, social) by reviewing 51 academic publications. 
Also by reviewing prior literature, Mcnutt~\etal~\cite{mcnutt2020surfacing} summarized a set of \textit{visualization mirages} (silent but significant failures in a visualization) and categorized them according to four stages of data analysis: curating, wrangling, visualizing, and reading.
Lo~\etal~\cite{lo2022misinformed} coded 1143 \textit{misinformative visualizations} collected from platforms such as Twitter and Google and proposed a taxonomy containing 74 design issues. The taxonomy was also framed according to the stages of analytics, including input, visualization design, plotting, perception, and interpretation.
Lisnic~\etal~\cite{lisnic2023misleading} analyzed 9958 misleading tweets that contain data visualizations about the COVID-19 pandemic and found that, apart from violating design rules, reasoning errors (\eg cherry-picking data, making causal inferences) constitute a significant part of misinformation about the pandemic.

While the aforementioned work provides constructive findings for the understanding of visualization design flaws, we believe more efforts can be made to facilitate the continuous development of this research direction.
For example, while our community has focused heavily on critiquing deliberate data distortion, the practical application of data visualization is more varied in form and intent. Therefore, it is valuable to keep investigating whether there are visualization design flaws that go beyond intentional deception and to incorporate more real-world examples in academic research.
Also, as most previous work has examined design flaws through reviewing academic literature or curating datasets by researchers themselves, more empirical studies based on firsthand flawed visualizations contributed and interpreted directly by users are desired.
Last, except for understanding \textit{what} constitutes a flawed visualization design, there is a lack of exploration into \textit{why} design flaws occur and \textit{how} to eliminate them.


Given the varied terms used in prior research such as ``deceptive visualization'', ``misleading visualization'', and ``visualization pitfalls'' without clear definitions, 
in this work, \textbf{we use the term \textit{design flaws} to encompass various problematic visualization designs, whether intentionally or unintentionally performed by designers, that hinder the effective presentation of data.} Building on this definition, this work aims to construct a task-driven taxonomy for visualization design flaws that differs from prior taxonomies in both its data source and the categories it proposes. Besides, we also conducted a series of focus groups to explore the root causes of these design flaws.


\subsection{Visualization for the General Public}


An increasing number of visualizations are being presented to and serving the general public rather than solely expert users. They are now found in diverse settings such as news media publications~\cite{shi2021communicating,lan2022negative}, advertisements~\cite{lan2021smile}, outdoor installations~\cite{lan2024affective,bressa2021s}, and personal devices~\cite{huang2014personal}. 
Given this trend, a lot of user studies have been conducted with the general public to understand their experience with data. 
Some findings have indeed shown that the study of the general public may provide valuable insights or reflections for the visualization community. For example, Cawthonand and Moere~\cite{cawthon2007effect} found that when presented with four visualizations (\eg treemap, sunburst diagram) that encode the same hierarchical data, people reported the least preference for the treemap, although it is often viewed as effective and efficient by researchers. 
Lan~\etal~\cite{lan2021smile} found from a crowdsourcing experiment that many participants reacted negatively to award-winning visualizations since the visual encodings were too novel or complex. Given such observations, more and more researchers agreed that we need to know more about such general users (also called non-expert users or novices sometimes~\cite{burns2023we}), such as understanding their criteria for evaluating visualization~\cite{quispel2014would,shukla2022negotiating} and their visualization literacy ~\cite{alper2017visualization,boy2014principled,lee2016vlat}.

In terms of methods to approach the general public, one strategy is to utilize crowdsourcing to examine how people perceive or interpret visualizations~\cite{borgo2018information}. 
Researchers have also used methods such as interviews and anthropomorphic studies (\eg ~\cite{peck2019data,lan2022negative}) to meet the general public, gathering their opinions and evaluations of visualizations.
Another method is to conduct content analysis based on the rich digital footprints left by the general public on the web. For example, Hullman~\etal~\cite{hullman2015content} collected comments posted on a visualization blog of The Economist and found that over one-third of the comments provided critiques for the visual presentation. 
Kauer~\etal~\cite{kauer2021public} analyzed 475 reactions to data visualizations on Reddit and categorized the reactions into ten categories (\eg observation, proposal). Our work follows this line of research and analyzes the visualizations and comments contributed by the general public on an online gallery to understand visualization design flaws through the lens of the general public.


\section{Design flaws characterized by the public}
\label{sec:space}

\begin{figure*}[!t]
 \centering
 \includegraphics[alt={An image of the taxonomy of 76 design flaws with frequency. There are 3 high-level categories, Misinformation, Uninformativeness, and Unsocialbility, with 10 subcategories, elaborated in the paragraphs.},width=0.85\textwidth]{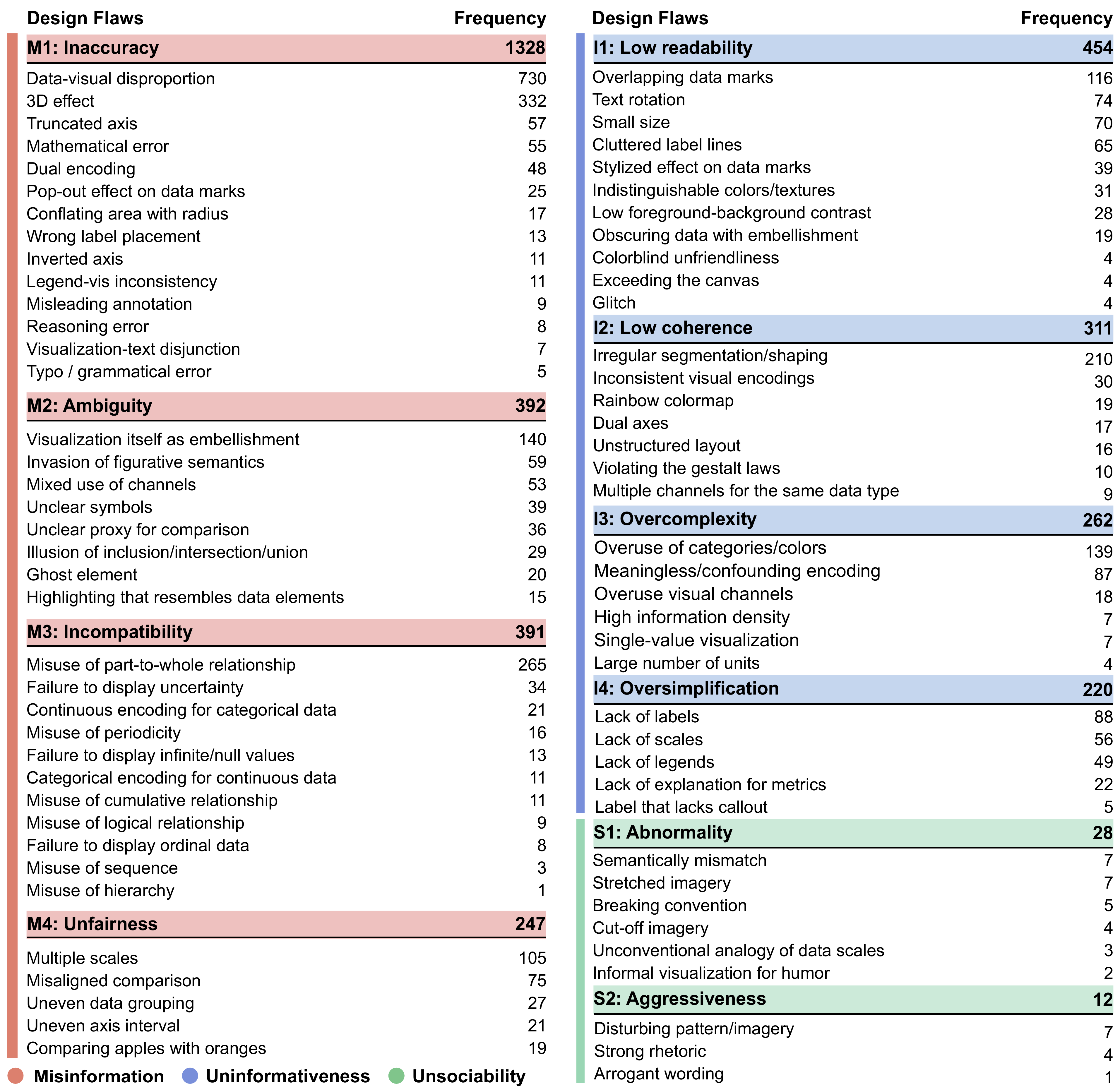}
 \caption{Visualization design flaws and their frequencies. More descriptions for each flaw can be browsed at \url{https://flawviz.github.io/}.}
 \label{fig:taxonomy}
 \vspace{-1em}
\end{figure*}

This section proposes a task-driven taxonomy of design flaws based on the analysis of flawed visualizations contributed by the general public.

\subsection{Methodology}

Below we introduce the corpus we collected and how we analyzed it.


\subsubsection{Corpus}

Our analysis was based on an online gallery called WTF Visualizations~\cite{wtf}, which was established in 2013 and has been showcasing flawed visualizations contributed by the public for ten years. 
One important feature of this galley is its democratic nature. Unlike many galleries that are constructed or led by professional editors, WTF Visualizations allows anyone to access and contribute to the collection, as well as write comments or explanations.
This fosters a more inclusive environment where individuals from different backgrounds can participate in and enables a crowdsourced approach to curating a wide variety of flawed visualization designs.
Therefore, we believe this gallery is both novel and of high quality for understanding the opinions of the general public, as well as the distribution of visualization design flaws in the wild.
We scraped all the visualizations as well as their metadata (\eg time of contribution, tags, user comments) from the gallery, resulting in a dataset containing 2297 images. Almost all the images have tags and comments contributed by their up-loaders to describe their visualization types (\eg bar chart, line chart) and design flaws (\eg 3D, rainbow). These tags and comments provided useful information and helped save the time of coding. 
Then, we performed data cleaning by excluding (1) duplicate images and (2) illustrations that contain no data. The final corpus comprises a total of 2227 data visualizations.


\subsubsection{Analysis}
Two authors were in charge of the coding process. In general, we adopted an open coding strategy and used the tags and comments provided by up-loaders as references.
As the number of visualizations was substantial, we followed previous studies~\cite{lo2022misinformed,lisnic2023misleading,lan2021smile} that also manually coded thousands of visualizations and divided the samples to be coded into a series of batches.
This method is good at enhancing coding efficiency while ensuring that codes reach saturation through incremental iterations.
Specifically, each coder first coded a batch (10\%, 220 visualizations) independently. Before coding, we familiarized ourselves with the naming of chart types (\eg ~\cite{borkin2013makes}) as well as the conceptualization of design flaws by previous literature (\eg ~\cite{lo2022misinformed,pandey2015deceptive}) to enhance the consistency of terminology. 
After the first round of independent coding, we met to compare our codes and discuss mismatches until reaching a Cohen's kappa > 0.7~\cite{neuendorf2017content}. Subsequently, we proceeded to code another batch of visualizations and refined the coding scheme through iterative discussions. After four rounds of coding, the codes showed clear convergence; the identification of new codes became less frequent, and our inter-coder agreement remained at a high level (> 0.7). Thereafter, we completed the coding of the remaining visualizations and initiated discussion only when encountering new codes. 
Finally, we organized the codes according to their associated design tasks. For instance, both the use of \textit{truncated axis} and \textit{dual encoding} violate the task of accuracy, leading to the delivery of incorrect messages. Consequently, they were categorized under the subcategory of \textit{inaccuracy} within the broader category of \textit{misinformation}.
As introduced earlier, adopting such a task-oriented analysis perspective assists in identifying common design tricks and strategies, revealing the functional impact of these design flaws on the visualization, and helping designers assess and refine their work using our taxonomy as heuristic tools~\cite{shi2021communicating,lan2021smile}.

\subsection{Taxonomy}

As shown in \cref{fig:taxonomy}, we identified 76 design flaws categorized into three high-level categories (\ie misinformation, uninformativenss, unsociability) and ten subcategories. Due to page constraints, we provide examples for a portion of the 76 design flaws, particularly those newly identified by this work compared to previous literature, in \cref{fig:example}. More detailed descriptions and examples for each design flaw can be found at \url{https://flawviz.github.io/}.

\subsubsection{Misinformation}
This category encompasses instances where visualization design delivers distorted or deceptive messages. Such design may be created inadvertently or deliberately (deliberate misinformation is also called disinformation). We identified four types of misinformation and 38 specific design flaws. The numbers in parentheses after each subcategory indicate the quantity of design flaws within it.

\begin{figure*}[t!]
 \centering
 \includegraphics[alt={26 example images containing design flaws across 3 high-level categories and 10 subcategories, more detailed explanations in paragraphs.},width=0.95\textwidth]{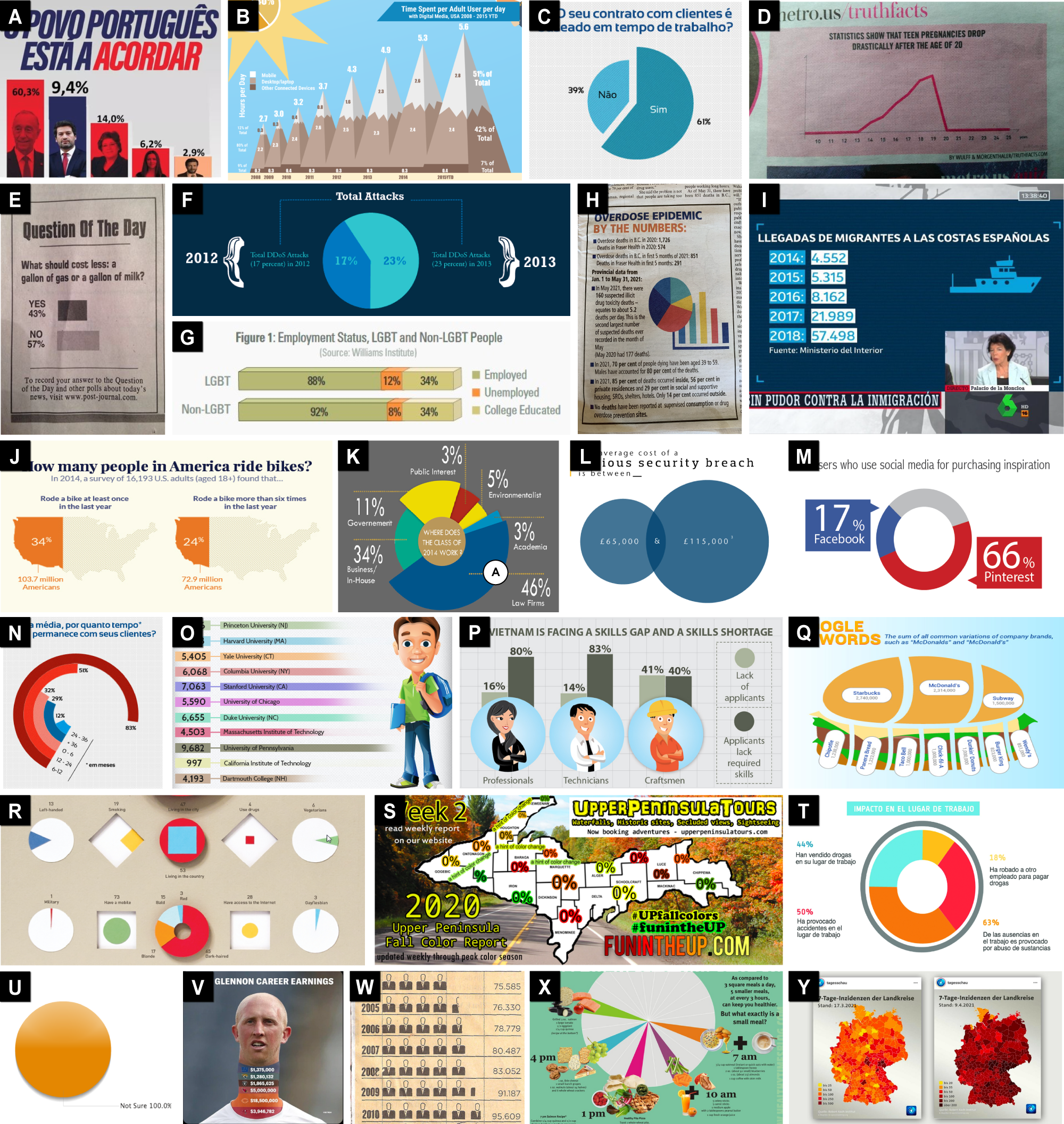}
 \caption{Examples of identified design flaws in data visualization, with indexes showing their corresponding categories.}
 \label{fig:example}
 \vspace{-1em}
\end{figure*}

\setul{0.3ex}{0.3ex}
\setulcolor{c1}

\ul{\textbf{M1: Inaccuracy (14).}} Inaccuracy occurs when the data points, labels, or other elements do not precisely represent the underlying data. The most observed flaws that belong to this subcategory include \textit{data-visual disproportion}, \textit{3D effect}, \textit{truncated axis}, \textit{mathematical error}, and \textit{dual encoding}.
For example, in \cref{fig:example} (A), the bar whose value is 9.4\% looks higher than the bar whose value is 14.0\%.
\cref{fig:example} (B) not only applies the 3D effect, which alters the perception of size, but also encodes data using dual channels (\ie height and width), thereby further exaggerating larger icons.
Other identified flaws in this subcategory include the \textit{pop-out effect on data marks}, \textit{conflating area with radius}, \textit{wrong label placement}, \textit{inverted axis}, \textit{legend-visualization inconsistency}, \textit{misleading annotation},, \textit{reasoning error} \textit{visualization-text disjunction}, and \textit{typo/grammatical error}. 
For example, \cref{fig:example} (C) applies the pop-out effect on the slice of 61\%, making it look much larger than it should be.
The visualization and text in \cref{fig:example} (E) are disconnected. The title asks: ``What should cost less: a gallon of gas or a gallon of milk?'' However, the visualization's outcome is ``yes'' and ``no''.
In \cref{fig:example} (D), we see a typical example of a reasoning error. The author concludes, based on the sharp decline in the line chart, that ``teen pregnancies drop drastically after the age of 20.'' However, the fact is that individuals over 20 are no longer counted as teens, which makes the rate appear to drop down to zero.

\ul{\textbf{M2: Ambiguity (8).}} Ambiguity occurs when a data visualization fails to clarify the mappings from data to visuals, making users uncertain of how to decode data.
The most salient flaws in this subcategory are \textit{visualization itself as embellishment}, 
\textit{invasion of figurative semantics},
\textit{mixed use of visual channels}, and  \textit{unclear symbols}.
For example, the charts in \cref{fig:example} (F) appear to be related to certain data mentioned in the text but do not actually represent any data.
The map-shaped pictograph in \cref{fig:example} (G) is used to show the proportion of bike riders in America. However, although the map serves solely as a decorative element, it has been misinterpreted to mean that only Western Americans ride bikes (\ie the semantics of the icon "invades" data interpretation). 
In \cref{fig:example} (H), the encodings of pie charts are mixed with Nightingale rose charts, creating ambiguity regarding whether to decode the angle or the length channel.
Other flaws include the \textit{unclear proxy for comparison}, textit{illusion of inclusion/intersection/union}, \textit{ghost element}, and \textit{highlighting that resembles data elements}.
For example, the circles in \cref{fig:example} (I) overlap as if they have a logical intersection (although in reality, there isn't one).
The donut chart in \cref{fig:example} (J) does not provide any explanation for the grey slice, leaving it as a ``ghost element''. In \cref{fig:example} (K), the background color is used to highlight the numbers, but this highlighting can easily be misinterpreted as a bar chart.

\ul{\textbf{M3: Incompatibility (11).}}
Incompatibility arises when data visualization is not appropriately tailored to input variables.
The dominant flaw in this subcategory is the \textit{misuse of part-to-whole relationship}.
For example, \cref{fig:example} (L) uses a pie chart to show temporal change from 2012 to 2013, and the two categories do not constitute a part-to-whole relationship. 
Other flaws in this subcategory include \textit{failure to show uncertainty}, \textit{continuous encoding for categorical data}, \textit{misuse of periodicity}, \textit{failure to display infinite/null values}, textit{categorical encoding for continuous data}, \textit{misuse of cumulative relationship}, \textit{misuse of logical relationships}, 
\textit{failure to display ordinal data},
\textit{misuse of sequence}, and \textit{misuse of hierarchy}. 
For example, In \cref{fig:example} (M), items like ``employed'', ``unemployed'', and ``college educated'' are added without a cumulative relationship. Additionally, many visualizations in our corpus misuse periodic tables by attributing properties or effects to elements without scientific basis or by drawing unwarranted connections.

\ul{\textbf{M4: Unfairness (5).} }
Unfairness presents different data points or categories following different standards or baselines. Relevant design flaws include \textit{multiple scales}, \textit{misaligned comparison}, \textit{uneven data grouping}, \textit{uneven axis interval}, and \textit{comparing apples with oranges}.
For example, \cref{fig:example} (N) uses multiple scales (circles with varied radii) to show proportions. Even though their proportions are similar, the outer circle looks much longer than the inner circle. 
In this subcategory, compared to other flaws that involve visual manipulation, \textit{comparing apples with oranges} is more special. For example, a typical instance we encountered is a chart that compares ``money'' with ``Obama''. These two variables lack comparability in terms of both units and semantics.

\subsubsection{Uninformativeness}

Uninformativeness refers to the absence of meaningful information, hindering users from processing data and gaining insights. Note that we also classify information overload as a form of uninformativeness, as it hides valuable insights within chaos. Specifically, we identified four subcategories of uninformativeness and 29 specific design flaws.

\setul{0.3ex}{0.3ex}
\setulcolor{c2}

\ul{\textbf{I1: Low readability (11).}}
Low readability means the data visualization is unreadable or illegible. This might stem from \textit{overlapping data marks}, \textit{text rotation}, \textit{small size}, \textit{cluttered label lines}, \textit{stylized effect on data marks}, \textit{indistinguishable colors/textures}, \textit{low foreground-background contrast}, \textit{obscuring data with embellishment}, \textit{colorblind unfriendliness}, \textit{exceeding the canvas}, or \textit{glitch}.
For example, in \cref{fig:example} (O), a semi-transparent style is applied to the bar chart, making it difficult for users to discern where the bars end. 
In \cref{fig:example} (P), decorative illustrations obscure the x-axis of the bar chart, preventing users from determining the exact length of the bars.

\ul{\textbf{I2: Low coherence (7).}} Low coherence indicates that the encodings of visualization lack consistent and unified criteria for users to comprehend the information they present. This could result from \textit{irregular segmentation/shaping}, and \textit{inconsistent visual encodings}, \textit{rainbow colormap}, \textit{dual axes}, \textit{unstructured layout}, \textit{violating the gestalt laws}, and \textit{multiple channels for the same data type}.
For example, in \cref{fig:example} (Q), the hot dog is divided into multiple irregularly-shaped segments, lacking a consistent visual structure, which makes interpreting the values inconvenient.
In \cref{fig:example} (R), multiple channels (\eg angle, circle area, and rectangle area) are used to present proportions, necessitating viewers to frequently switch channels to interpret the charts.
The textual labels in \cref{fig:example} (S) are misplaced and do not align with the slides they correspond to, thus violating the Gestalt law of proximity.

\ul{\textbf{I3: Overcomplexity (6).}} 
Overcomplexity happens when a data visualization becomes overly intricate and convoluted. 
Relevant flaws include \textit{overuse of categories/colors}, \textit{meaningless \& confounding encoding}, \textit{overuse of visual channels}, \textit{high information density}, \textit{single-value visualization}, and \textit{a large number of units}. 
For example, the application of both coloring and sizing to every "0\%" instance in \cref{fig:example} (T) is redundant, as these two visual channels fail to convey meaningful information.
\cref{fig:example} (U) visualizes a single value (\ie, 100\%) as a large circle, which is unnecessary.

\ul{\textbf{I4: Oversimplification (5).} }
Oversimplification, as the opposite of overcomplexity, occurs when a data visualization misses essential components, such as the \textit{lack of labels}, \textit{lack of scales}, \textit{lack of legends}, \textit{lack of explanation for metrics}, and \textit{label that lacks callout}. For example, Figure \ref{fig:example} (V) showcases a bar chart devoid of a y-axis and scale, rendering it impossible for users to understand what the bars represent or their corresponding values. 

\setul{0.3ex}{0.3ex}
\setulcolor{c3}

\subsubsection{Unsociability}

Unsociability happens when a visualization makes people feel uncomfortable, offended, or socially awkward. We identified two subcategories of unsociability and 9 specific design flaws.


\ul{\textbf{S1: Abnormality (6).} }
Abnormality signifies instances where the design choices of the data visualization diverge significantly from standard practices and social norms. This can create confusion and hinder intuitive understanding, and relevant flaws include the \textit{semantically mismatch}, \textit{stretched imagery}, \textit{breaking convention}, \textit{cut-off imagery}, \textit{unconventional analogy of data scales} and \textit{informal visualization for humor}. 
For instance, \cref{fig:example} (W) stretches a person's neck to present a bar chart, which creates peculiar imagery. \cref{fig:example} (X) visualizes the time taken to consume food using a pie chart, but the placement of times does not conform to the traditional clock layout, leading to discomfort. \cref{fig:example} (Y) uses human-shaped icons to represent data. However, with the introduction of fractional values, the ``human beings'' are abruptly truncated, which can make people feel uneasy.

\ul{\textbf{S2: Aggressiveness (3).}}
Aggressiveness creates a negative user experience by employing overly hostile or forceful visual elements, such as the \textit{disturbing imagery/patterns}, \textit{strong rhetoric}, and \textit{arrogant wording}. 
For example, \cref{fig:example} (Z) presents two versions of a visualization illustrating the distribution of virus infection. The version on the right was criticized by the uploader for creating overly serious rhetoric by modifying the range and intervals of the color palette.
Regarding \textit{arrogant wording}, we encountered a visualization concerning social disparity whose title stated, ``Poor people need to primarily change their approach to life,'' in a condescending manner. This visualization angered a large number of users and was shared extensively.

\subsection{Observations}

Below we discuss our observations on the taxonomy from two aspects: the distribution of the design flaws within the taxonomy and how it differs from previous taxonomies.

\subsubsection{Distribution of design flaws}

As summarized in \cref{fig:taxonomy}, the design flaws showing high frequency (N>100) in our taxonomy are \textit{data-visual disproportion} (730), \textit{3D effect} (332), \textit{misuse of part-to-whole relationship} (265), \textit{irregular segmentation/shaping} (210), \textit{visualization itself as embellishment} (140), \textit{overuse of categories/colors} (139), \textit{overlapping data marks} (116), and \textit{multiple scales} (105). 
In terms of the ten subcategories, \textit{inaccuracy} (1328) has the highest frequency, and the frequencies of \textit{low readability} (454), \textit{incompatibility} (392), \textit{ambiguity} (391), and \textit{low efficiency} (315) are also high.
An interesting finding is that the aforementioned high-frequency design flaws distribute relatively evenly across different categories, confirming that design flaws can arise from various aspects and be associated with diverse tasks.

We also calculated the distribution of chart types. Notably, the pie chart (394, 18\%), donut chart (345, 15\%), and bar chart (343, 15\%) emerge as the most prevalent chart types with reported design flaws, 
followed by the ISOTYPE chart (296, 13\%), proportional area chart (130, 6\%), line chart (123, 6\%), circular bar chart (88, 4\%), stacked bar chart (65, 3\%), progress bar chart (61, 3\%), nightingale rose chart (59, 3\%), and schematic diagram (57, 3\%). 
For pie charts and donut charts, the most dominant flaws are similar, including the \textit{3D effect} (139 and 32, respectively), \textit{misuse of part-to-whole relationship} (111, 89), and \textit{data-visual disproportion} (96, 142).
For bar charts, the most dominant flaws are the \textit{data-visual disproportion} (182), \textit{3D effect} (74), and \textit{truncated axis} (21).
For ISOTYPE charts, the most dominant flaws are the \textit{data-visual disproportion} (122), \textit{irregular segmentation/shaping} (104), and \textit{invasion of figurative semantics} (37).
In short, most of these flawed visualizations are based on the bar, circle, or pictorial mark. Data-visual disproportion is a universal problem across nearly all chart types, while other design flaws are more chart-specific.



\subsubsection{Comparison with previous taxonomies}

We compared our taxonomy in \cref{fig:taxonomy} with existing taxonomies~\cite{lisnic2023misleading,lo2022misinformed,mcnutt2020surfacing} and derived three main findings:

\textbf{A set of new design flaws have been identified.}
Although some design flaws have been identified by all taxonomies, including \textit{truncated axis}, \textit{conflating area with radius}, \textit{inverted axis}, \textit{dual axes}, and \textit{wrong reasoning}, this work identified 36 new design flaws that have not been explicitly characterized by previous taxonomies (marked as \includegraphics[height=0.7em]{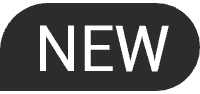} on our website), such as \textit{highlighting that resembles data elements}, \textit{invasion of figurative semantics}, \textit{illusion of inclusion/intersection/union}, \textit{ghost element}, and \textit{violating the gestalt laws}. Meanwhile, we also identified more specific design flaws previously categorized as broad terms \textit{overplotting}, \textit{unclear encoding}, and \textit{inappropriate encoding}~\cite{lo2022misinformed,lisnic2023misleading}.
Particularly, nearly all design flaws in the category of unsociability have not been characterized by previous research. This suggests that visualization serves not only as a scientific means of transforming data but also as a medium intertwined with social factors. Furthermore, this discovery underscores the importance of conducting further empirical research, as design flaws in visualizations may be more nuanced and extensive than anticipated by traditional visualization theories.

\textbf{Data-visual disproportion is pervasive but lacks investigation.} Different from previous taxonomies, \textit{data-visual disproportion} is the design flaw with the highest frequency in our study. Notably, we found this flaw to be difficult to explain using existing visualization theories. As shown in \cref{fig:example} (A), many instances of these flawed visualizations lack axes or referential scales, making it impossible to determine whether the authors truncated the axes, misunderstood certain data components, or simply drew the graphics hastily and irresponsibly. We found such flaws to be abundant in the wild, troubling many users, yet they still lack theoretical attention from academia.

\textbf{Embellishment's role is more intricate than previously explored.} Although in this work we did not treat embellishment as an inherent flaw, we did find rich instances where embellishment obstructs data presentation. Such flaws span across \textit{misinformation}, \textit{uninformativeness}, and \textit{unsociability}, and many of them have received scant attention in both previous taxonomies and user studies~\cite{bateman2010useful,borgo2012empirical,andry2021interpreting,haroz2015isotype,alebri2023embellishments}. For instance, previous studies mainly examined cases where embellishment is placed adjacent to visualizations or created as ISOTYPE. However, our study revealed a more diverse range of applications for embellishment. It may partially overlay data marks, serve as stylistic filters superimposed on data marks, or become deeply intertwined with data through visual metaphors, thereby creating complex situations that compromise data to accommodate embellishment (\eg \cref{fig:example} (O, Q)). Moreover, we found that the semantics of embellishment itself can impact visualization, such as causing distraction or discomfort (\eg \cref{fig:example} (J, W)).

\section{Inferring why design flaws happen}
\label{sec:focus}

After characterizing the landscape of visualization design flaws in the eyes of the public, we further explore \textit{why} these flaws happen.

\subsection{Methodology}

Although the most ideal method to understand how the design flaws arose is to locate their original authors and inquire with them, this approach lacks practicality for our work. The primary reason is that the quantity of visualizations is large in our corpus, and most works lack information about their authors. Additionally, these images themselves are not contributed by their authors, but by uploaders who have critical intentions towards their design. Thus, contacting the original authors may lead to significant obstacles. Therefore, we opted for a compromise by inviting a series of participants with experience in visualization design to observe and discuss these flaws. Inspired by the idea of elicitation~\cite{hogan2015elicitation} (\ie interacting with human subjects to elicit information from them), we aimed to stimulate their recollection and association with their own relevant experiences in visualization design, especially situations where they have knowingly or unknowingly made similar flaws, in order to understand why these design flaws occur.

Specifically, we conducted a series of focus group studies. Focus group is a frequently adopted research method to understand people in-depth and gain qualitative data. In comparison to other elicitation methods such as interviews, the focus group emphasizes the interactive discussions among a diverse group of participants, allowing for the emergence of multifaceted perspectives, spontaneous exchange of ideas, and the identification of tensions or controversies~\cite{krueger2014focus}, which better aligns with our research goal.

\subsubsection{Stimuli}
 
As it is unrealistic to use all the 2227 images in our corpus as stimuli in focus group studies, we referred to previous work and selected a subset of images as stimuli~\cite{borkin2013makes}. When selecting the subset, we considered three criteria. First, the stimuli should cover all the visualization types and design flaws identified in the original corpus. Second, the distribution of visualization types and design flaws should match that of the original corpus as much as possible. Third, the stimuli should be of high resolution.
Following these criteria, we selected a collection of 129 representative visualizations as stimuli. 
For offline focus groups, the stimuli were printed in A4 size. 


\subsubsection{Participants}
To recruit participants for the focus groups, we posted open invitation posters on social media platforms and chat groups centered around visualization subjects. 
When recruiting the participants, we referred to the guidelines for focus group research~\cite{krueger2014focus}. For example, to guarantee effective discussion among participants, the size of a focus group should not be too small or large (a suggested size is 5-10 people per group). Besides, the backgrounds of the participants should be diverse and there should be no power differentials among the participants. In addition, to gain as many insights as possible for our research question (\ie why design flaws happen), we expected that the participants should at least have some experience with creating data visualization. 
With these in mind, we created a questionnaire to collect information from the applicants, including three open-ended questions (\ie gender, age, job/major) and one rating question that assessed expertise with data visualization (5-point scale ranging from ``Novice'' to ``Expert'').
At last, we recruited 29 participants (17 females, 12 males) aged from 21 to 52 (\textit{M} = 29.24, \textit{SD} = 6.83). Their jobs/majors were diverse: 32\% were data analysts, 14\% were visual designers, 8\% were product managers, 14\% were developers, and 32\% were students with different majors (\eg journalism, design, geography). All the participants had experience with visualization design (Expert: 4\%, Proficient: 21\%, Competent: 46\%, Beginner: 29\%, Novice: 0\%).
The participants were assigned to five distinct focus groups and each focus group involved 5-8 participants (see more details in \cref{tab:participants}). 
All the participants were compensated with a \$20 gift card.

\renewcommand{\arraystretch}{1}
\begin{table}[t!]
\centering
\fontsize{7.8}{8.5}\selectfont
\caption{Information of the 29 participants of our focus group studies.}
\begin{adjustbox}{width=\columnwidth}\begin{tabular}{llllll}
\toprule
Group & ID &	Gender &	Age &	Job/Major &	Expertise \\
\midrule
G1	&	P1	&	F	&	25	&	Visualization Student	&	Competent	\\
G1	&	P2	&	F	&	22	&	Business Analysis Student	&	Beginner	\\
G1	&	P3	&	F	&	33	&	Visualization Developer	&	Proficient	\\
G1	&	P4	&	F	&	30	&	Data Analyst	&	Competent	\\
G1	&	P5	&	M	&	32	&	Data Analyst	&	Competent	\\
G1	&	P6	&	F	&	25	&	Designer	&	Beginner	\\
G1	&	P7	&	M	&	31	&	Project Manager	&	Beginner	\\
G1	&	P8	&	F	&	25	&	Visualization Student	&	Proficient	\\
G2	&	P9	&	F	&	26	&	Visualization Student	&	Proficient	\\
G2	&	P10	&	F	&	28	&	Data Consultant	&	Beginner	\\
G2	&	P11	&	M	&	28	&	Front-end Developer	&	Competent	\\
G2	&	P12	&	M	&	27	&	Business Analyst	&	Competent	\\
G2	&	P13	&	F	&	33	&	Product Manager	&	Competent	\\
G3	&	P14	&	F	&	29	&	Designer	&	Proficient	\\
G3	&	P15	&	F	&	37	&	Data Analyst	&	Beginner	\\
G3	&	P16	&	M	&	33	&	Front-end Developer	&	Competent	\\
G3	&	P17	&	M	&	22	&	Design Student	&	Competent	\\
G3	&	P18	&	F	&	30	&	Data Consultant	&	Competent	\\
G3	&	P19	&	F	&	27	&	Visualization Developer	&	Competent	\\
G4	&	P20	&	F	&	27	&	Designer	&	Proficient	\\
G4	&	P21	&	M	&	26	&	Business Analyst	&	Competent	\\
G4	&	P22	&	F	&	21	&	Computer Science Student	&	Proficient	\\
G4	&	P23	&	M	&	40	&	Data Analyst	&	Beginner	\\
G4	&	P24	&	M	&	33	&	Data Analyst	&	Competent	\\
G5	&	P25	&	M	&	40	&	Data Analyst, Designer	&	Expert	\\
G5	&	P26	&	F	&	21	&	Data Journalism Student	&	Beginner	\\
G5	&	P27	&	M	&	21	&	Design student	&	Beginner	\\
G5	&	P28	&	F	&	24	&	Cartography Design Student	&	Competent	\\
G5	&	P29	&	M	&	52	&	Data Analyst	&	Competent	\\
\bottomrule
\end{tabular}
\end{adjustbox}
\label{tab:participants}
\vspace{-2em}
\end{table}

\subsubsection{Procedure}

The five focus groups were conducted separately. Three were conducted online and two were conducted offline. For the online studies, we used video conferencing software to make the participants meet and discuss with each other and collaborative design software to assign each participant a ``seat'' on the canvas and present them with stimuli on the corresponding ``desk''. The participants can annotate their stimuli (\eg add sticky notes, write down texts) as if offline. All the focus groups followed an identical procedure, including an introduction stage, a discussion stage, and a conclusion stage. The procedure was designed following the methodology suggested by Krueger~\cite{krueger2014focus} to ensure the participants feel comfortable and willing to express opinions while collecting high-quality data for research.

In the introduction stage, we began by asking the participants to introduce themselves one by one to break the ice. Next, we explained the theme of our research and provided essential background information (\eg the chart junk debate) to the participants. This stage lasted about 20 minutes.
Note that since the term \textit{chart junk} can be interpreted in various ways~\cite{parsons2020data} and some researchers have even suggested rephrasing it~\cite{akbaba2021manifesto}, we were cautious about using it during the focus group study to avoid restricting or misleading participants' thoughts and expressions. Therefore, apart from mentioning the chart junk debate in the introduction stage, we used the term \textit{design flaw} instead (with its definition provided as defined in \cref{sec:formulation}) throughout the study to guide discussions.
In the discussion stage, each participant was assigned 4-5 images randomly chosen from our stimuli. They first viewed the images independently for about 10 minutes with the goal of identifying design flaws, free to annotate the images with pens or sticky notes.
Next, one of the authors moderated the group discussion. The discussion was centered on two key questions, namely ``What design flaws have you found?'', and ``What do you think caused this fault?''. The moderator also asked follow-up questions such as ``Could you please explain more about this point?'' and ``Does anyone agree or disagree?'' if the participants did not express themselves clearly or to insert more dynamics into the discussion. The discussion stage lasted about 70 minutes and another researcher was in charge of taking notes of the discussion.
In the conclusion stage, we summarized the findings from the focus group and encouraged free discussion about how to combat such flaws. This stage lasted approximately 30 minutes.
The whole study lasted about 2 hours and was audio-recorded with the participants' consent.


\subsection{Results}


We transcribed the audio recordings and then coded the texts with the goal of identifying the answers to our research question using thematic analysis~\cite{braun2012thematic}. Two authors were in charge of the coding process. First, we coded the transcriptions independently and marked any sentences related to our research questions. Then, we read through all the marked sentences, generated codes from the sentences, and grouped similar codes as higher-level themes. 
After independent coding, we met and compared our codes and discussed mismatches until reaching 100\% agreement.
At last, we in total identified seven main causes of flawed visualization design (R1-R7) and split them into 15 specific points. \\

\begin{wrapfigure}[3]{l}{0.035\textwidth}
\centering
\vspace*{-12pt}
\includegraphics[width=0.05\textwidth]{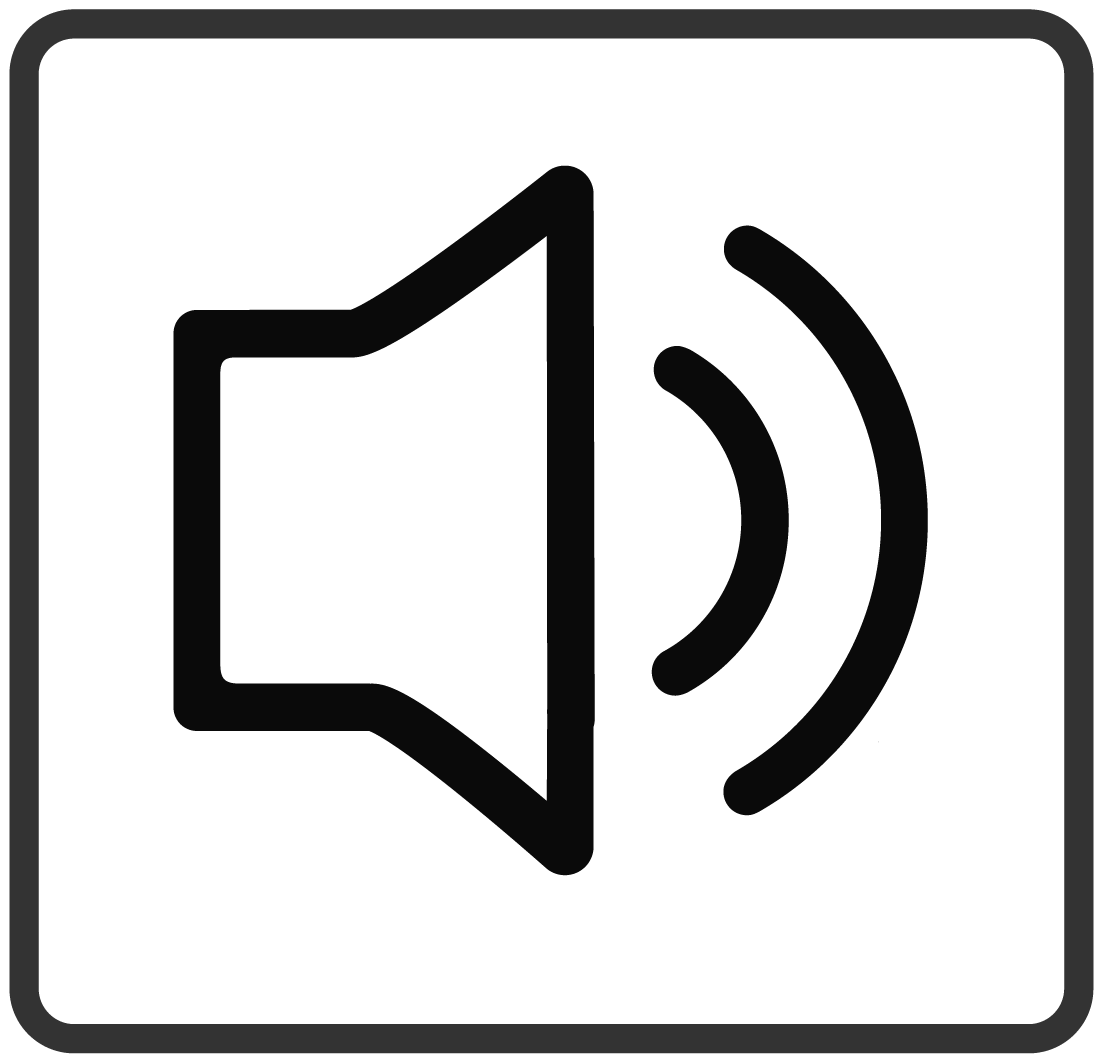}
\end{wrapfigure}
\noindent
\textbf{R1: Achieving communication intents.}
Ten participants mentioned that visualization design can be manipulated to achieve intents such as persuasion and boosting traffic.

\underline{R1.1 Conveying pre-defined opinions or values.} Sometimes, the designer of a visualization already knows the message he/she wants to communicate before the visualization is done. For example, when viewing a visualization for a political election, P15 commented that ``\textit{politicians always want their own statistics looks beautiful and their competitors' approval ratings as low as possible}''. For another visualization about market growth, P16 said, ``\textit{The designer added a misleading annotation on this chart to emphasize the market is expanding, maybe to strengthen customers' confidence or to attract more investment.}''
P3, who once worked as a journalist, said that ``\textit{I used to make data stories. I know the media will have a certain left/right-wing bias, and that's reflected in the visualizations they produce.}'' 

\underline{R1.2 Attracting attention and clicks.}
Flawed design can also be made to attract or engage viewers. For example, P27 thought that lavish visual effects such as 3D and abundance of color could ``\textit{make the visualization stand out from other content and grab my eyes immediately.}'' When expressing opinions about a visualization whose embellishment has obstructed data, P23 commented, ``\textit{At first glance, it seems to be very eye-catching, and the designer has achieved his/her purpose...viewers do not need to delve into the data inside...having a first impression is enough.}''

\underline{R1.3 Creating a sense of ``science''.}
When discussing why some designers would add a non-sense visualization that encodes no data to the image, P5 pointed out that ``\textit{visualization is a good medium for creating a scientific atmosphere...for example, in a presentation, even if you did not do any statistics, you want to add a chart on the slides to show you are doing things seriously}''. P20 added that ``\textit{it seems that putting a visualization aside your texts can make your arguments seemingly based on data and thus trustworthy.}'' \\

\begin{wrapfigure}[3]{l}{0.035\textwidth}
\centering
\vspace*{-12pt}
\includegraphics[width=0.05\textwidth]{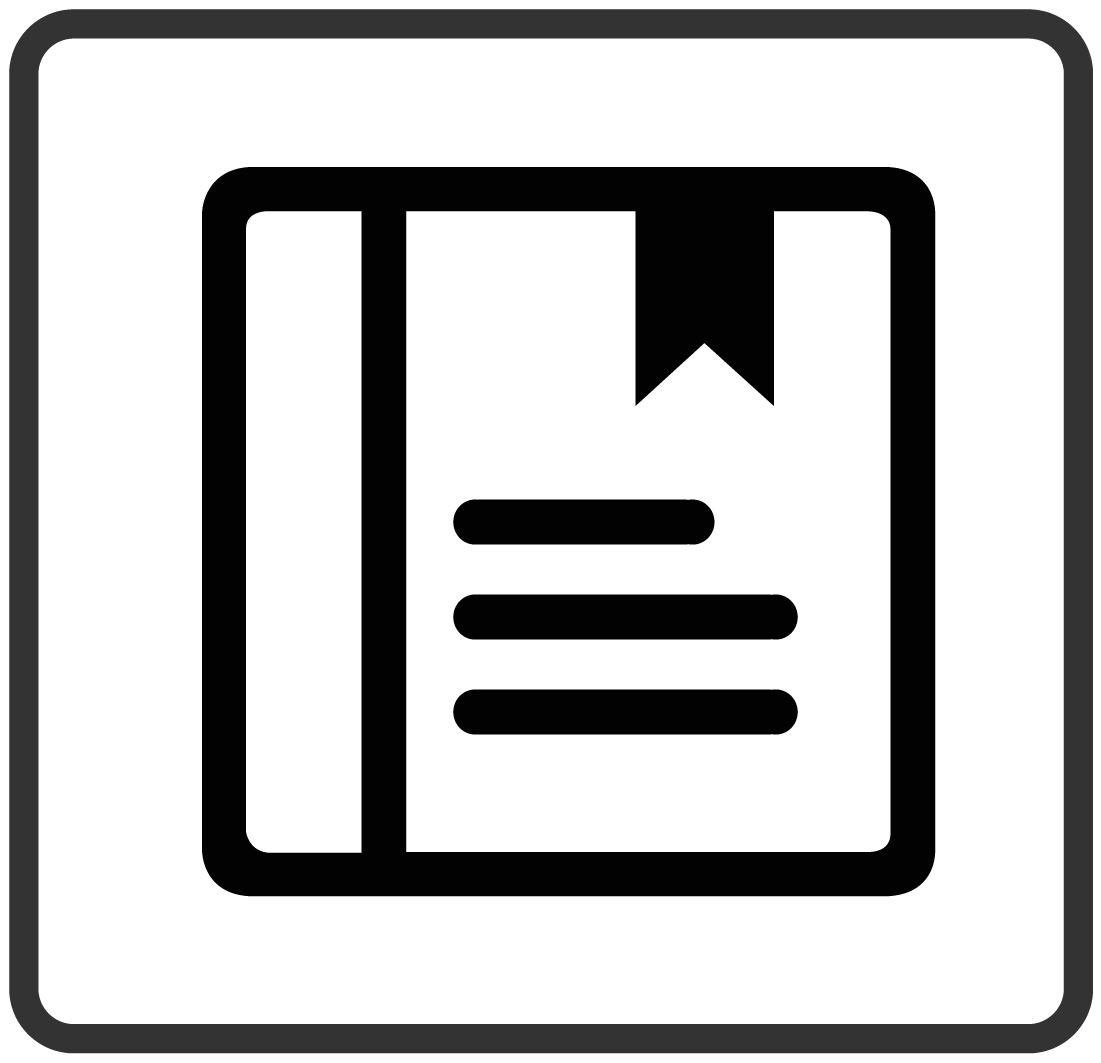}
\end{wrapfigure}
\noindent
\textbf{R2: Lack of literacy and expertise.}
Seven participants thought visualization design flaws could be made by designers who are not proficient in handling data or visuals.

\underline{R2.1 Lack of data science/visual design literacy.}
P9 thought that the overuse of the rainbow palette may be attributed to ``\textit{the designer's limited knowledge about color and the inappropriate use of color tools.}''
P16 agreed that ``\textit{the flaws may not be intentionally made...perhaps the designer is not that familiar with this type of data visualization.}''
P3 talked about the interdisciplinary nature of data visualization and said, ``\textit{Not all creators of data visualization have gone through systematic education or training of data science. Many of them are graphic designers or journalists before, and they are more likely to make mistakes in data computation or graph choice.}''

\underline{R2.2 Too ambitious to tell a big story.}
When discussing visualization with messy layouts, P8 said, ``\textit{the designer could have used a series of simple charts to convey this information, but instead, they chose to cram so much information into one graph; with such a large goal and insufficient ability, it resulted in this disaster.}''
P16 recalled his own experience and agreed that ``\textit{I encounter such situations in my work as well. You may want to tell a big story in one single graph, aiming to do something impressive. However, when you fail to strike a balance between the density and clarity of information, you create a chart junk.}''  \\

\begin{wrapfigure}[3]{l}{0.035\textwidth}
\centering
\vspace*{-12pt}
\includegraphics[width=0.05\textwidth]{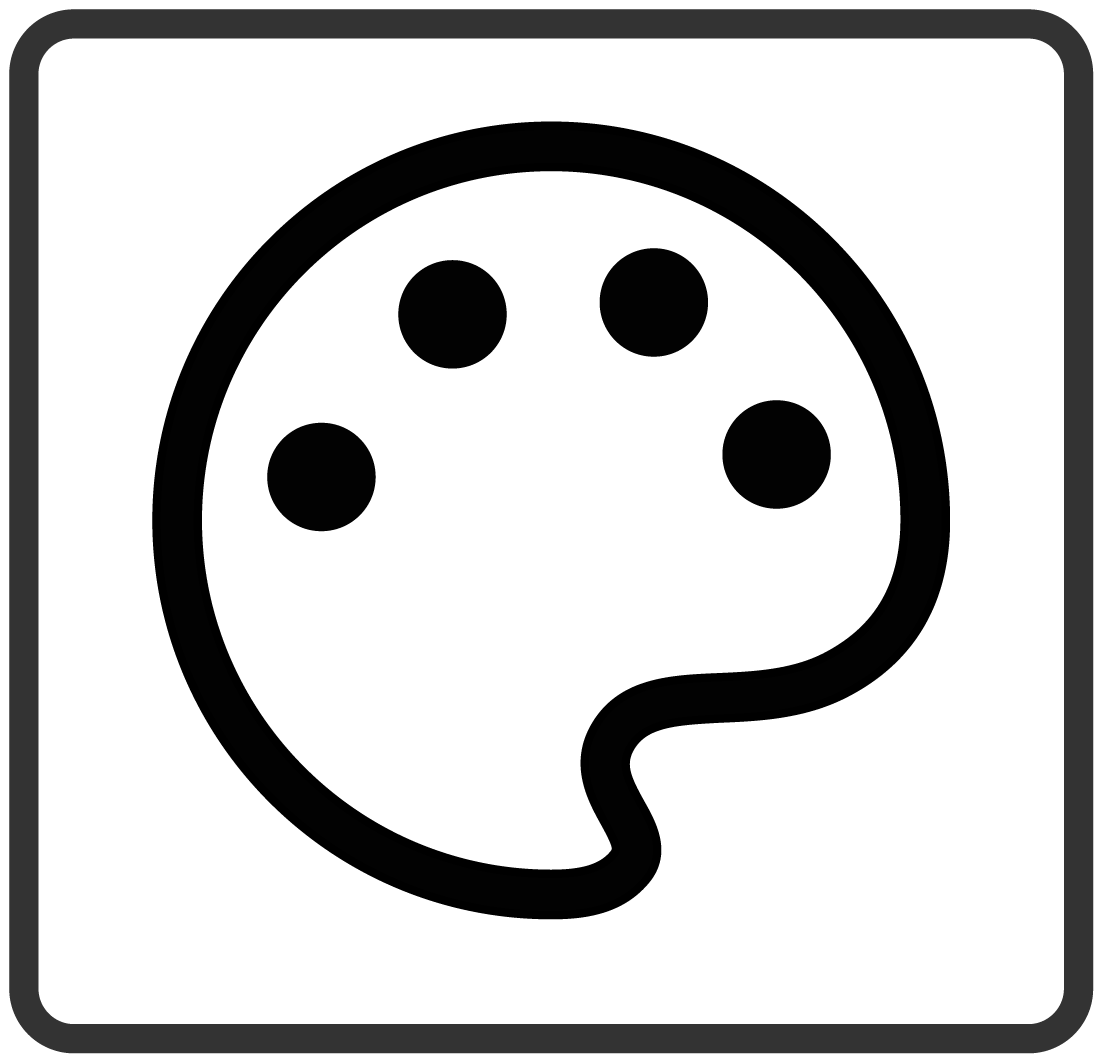}
\end{wrapfigure}
\noindent
\textbf{R3: Pursuit of aesthetics.}
Six participants mentioned that flawed visualization design can be made because the designer outweighs beauty over science.

\underline{R3.1 Beauty comes first.}
When viewing a donut chart that wrongly shows part-to-whole relationship, P8 thought, ``\textit{the circular shape of the donut is charming...the designer probably decided to use the donut first, and then fill data into it.}'' As a visual designer, P20 said ``\textit{I indeed put a lot of effort into thinking and modifying how a visualization looks like...aesthetics matter to me and sometimes visual elements are chosen before encoded with data.}'' Similarly, P14 thought, ``\textit{you can make a correct chart by following the standards on a visualization textbook, but this does not necessarily lead to a good-looking result.}''

\underline{R3.2 Exploring novel representations.}
When viewing a visualization that utilizes complex and unconventional channels to encode data, P14 mentioned, ``\textit{If you browse the gallery of Tableau Public, you can also see many visualizations like this one. They are actually more close to artworks. Unlike data visualization for business analysis, these works are exploring the boundaries of visual representations, or I would say they are researching some cutting-edge visualization methods so that clarity is not the primary consideration.}''  \\

\begin{wrapfigure}[3]{l}{0.035\textwidth}
\centering
\vspace*{-12pt}
\includegraphics[width=0.05\textwidth]{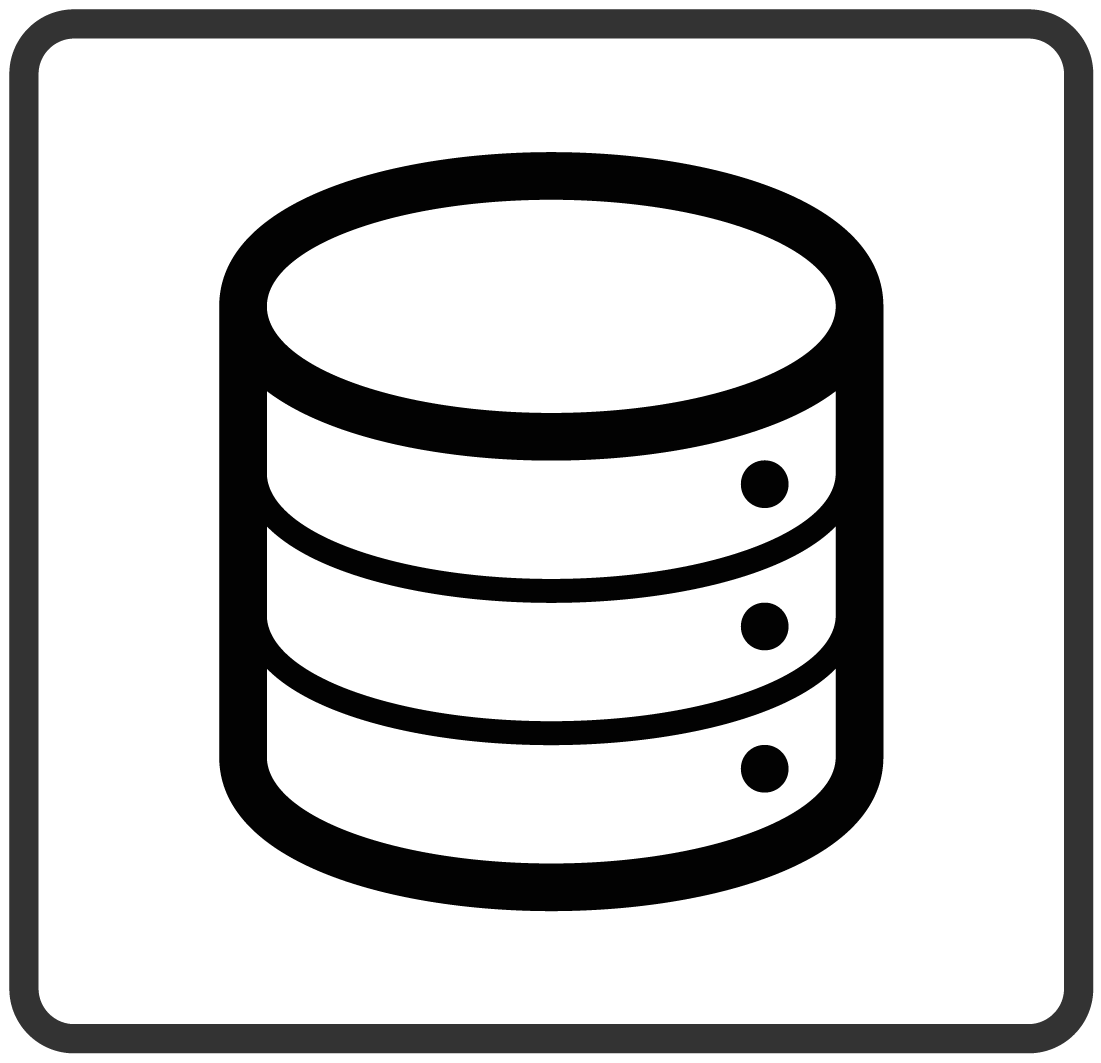}
\end{wrapfigure}
\noindent
\textbf{R4: Data that is difficult to handle.}
Six participants mentioned that design flaws may be inevitable when the dataset is too complex or lacks pre-processing.

\underline{R4.1 Data is complex and multidimensional.}
When viewing a map, P7 found that the same color is used to represent different categories. He said, ``\textit{this design is confusing of course. But there are so many categories, if they give one distinct color to a category, that would be even worse.}'' 
P19 complemented, ``\textit{Sometimes there are too many dimensions in the data, and you really don't know what to do. Not all designers are familiar with advanced charts like parallel coordinates. Even if they know, they may hesitate to use them because most people still have limited knowledge about these professional, or I might say academic charts.}''

\underline{R4.2 Inadequate data pre-processing.}
As said by P22, ``\textit{when the range of data is particularly large, directly visualizing it can result in overlapping or invisible data points. In such cases, it is advisable to preprocess the data by taking logarithms, for example.}''
P23 thought that the glitches in some visualizations may be caused by missing values (``\textit{The pixel missing here may be caused by missing values, and the author should make some cleaning or provide additional explanations.}'').  \\

\begin{wrapfigure}[3]{l}{0.035\textwidth}
\centering
\vspace*{-12pt}
\includegraphics[width=0.05\textwidth]{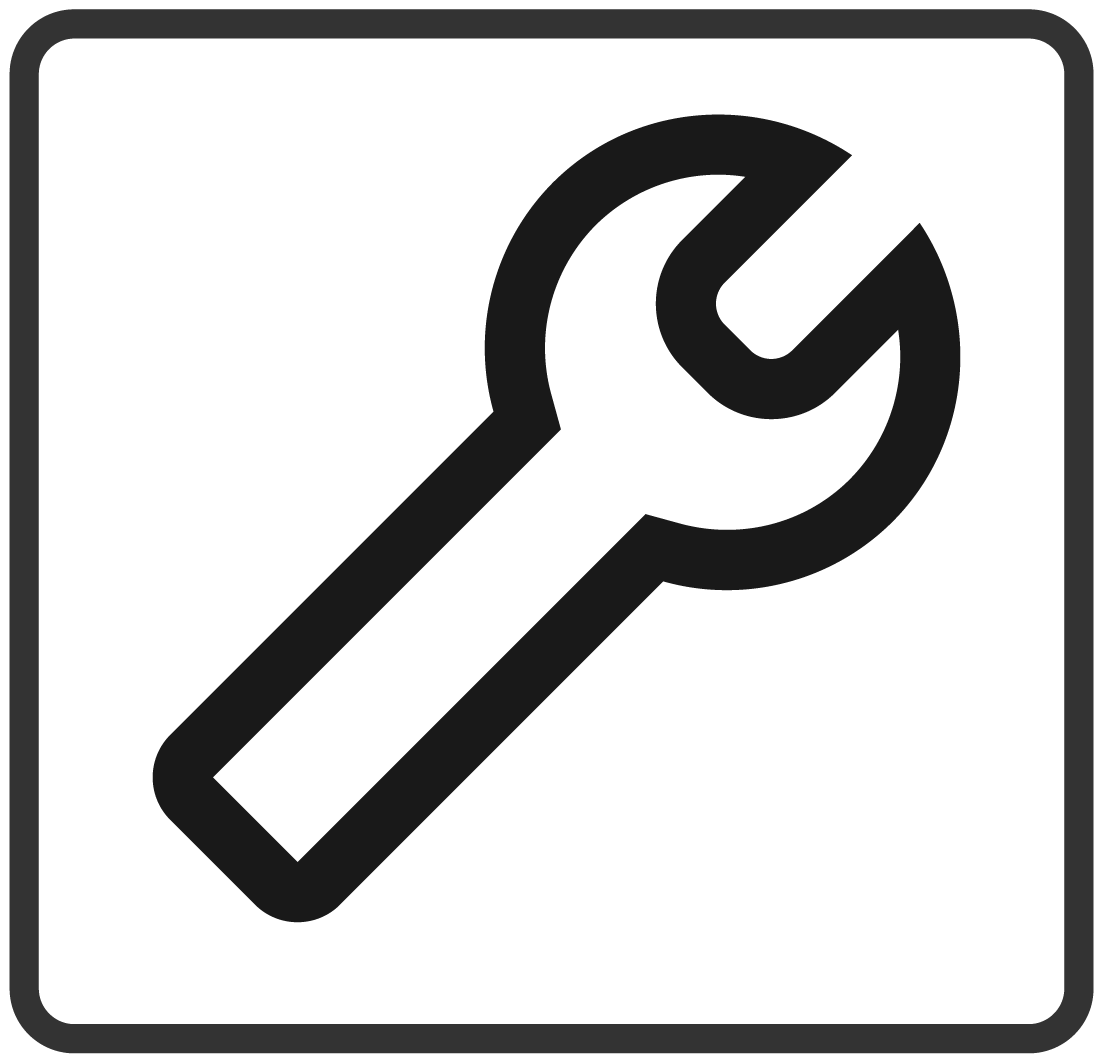}
\end{wrapfigure}
\noindent
\textbf{R5: Limitations of tools.}
Five participants mentioned that the occurrence of some design flaws may be attributed to the limitations of tools.

\underline{R5.1 Lack of data mapping functionality.}
Some design tools are not specialized for data visualization and cannot bind data. When viewing a visualization where data does not proportionally map to visual channels, P17 speculated that ``\textit{it might have used a PowerPoint template directly, which cannot bind data, so the creator just modified the text.}''
P8 noticed that more than one visualization she examined had issues such as misaligned elements and inaccurate data. She believed that ``\textit{this is because the author did not use professional visualization tools to create the charts, but rather drew them by hand...look at this pie chart, its center is not even in the middle.}'' 

\underline{R5.2 Features/bugs of tools.}
P3 commented that ``\textit{tools can produce errors. I've encountered situations when my data went wrong while using D3.js. If you are not familiar with the underlying computational mechanism, it can be difficult to resolve them.}''
P20, as a visual designer, explained using the example of creating a bubble chart: ``\textit{When you draw a circle, design software like Adobe Illustrator directly maps the input value to the area of the circle. However, in code libraries like D3, the input value should be the radius of the circle. This means that when using different tools, even with the same input values, the resulting circle areas may differ several times.}''  \\

\begin{wrapfigure}[3]{l}{0.035\textwidth}
\centering
\vspace*{-12pt}
\includegraphics[width=0.05\textwidth]{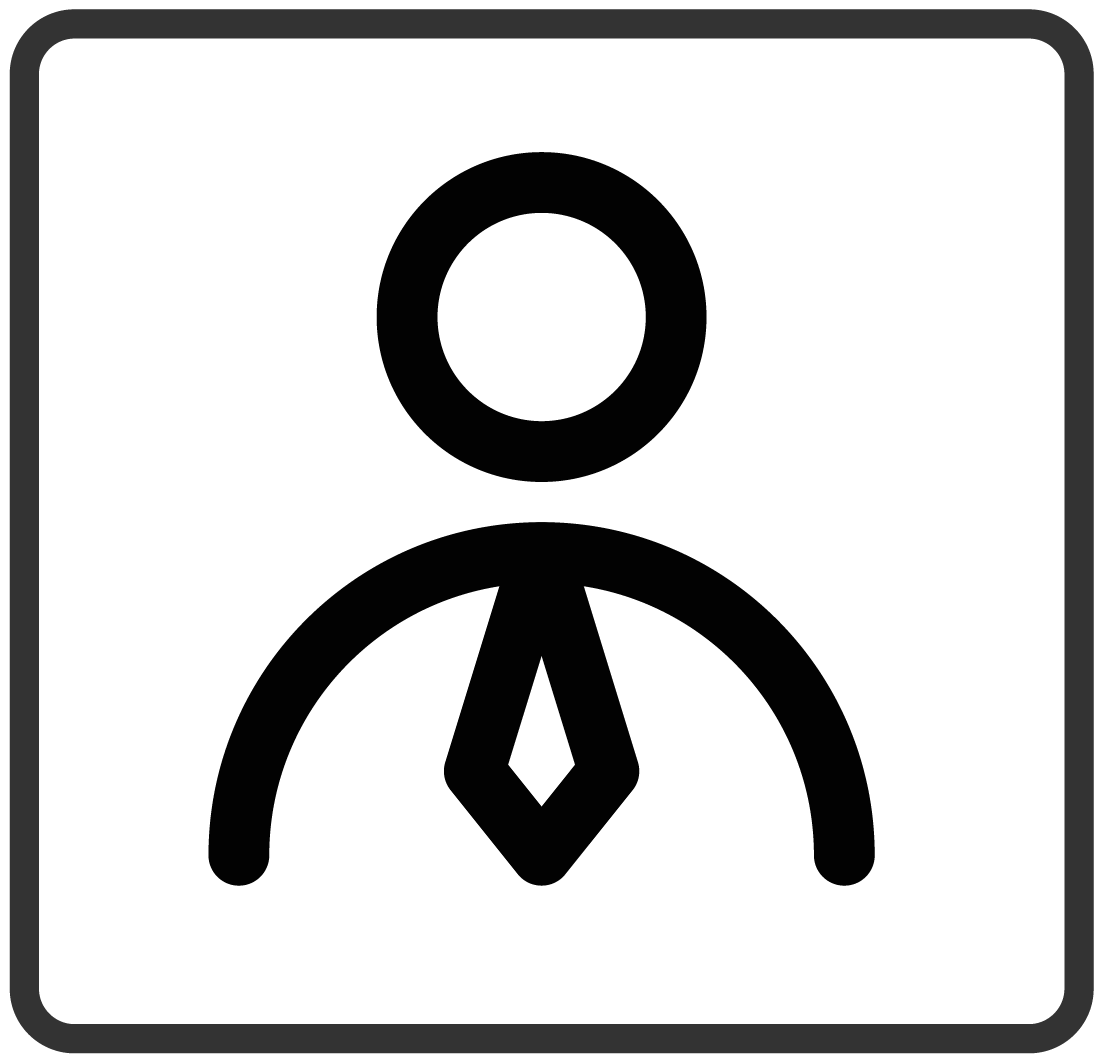}
\end{wrapfigure}
\noindent
\textbf{R6: Compromise with clients.}
Three participants mentioned that some design flaws can be caused by the unprofessional or unrealistic needs required by clients.

\underline{R6.1 Clients are non-professionals.}
P5, as a data analyst who often collaborates with business suppliers and governments, said, ``\textit{Most clients are non-experts in data visualization, but they pay, and they have the right to decide whether your work is qualified. This will certainly distort your work. Frequently, I can only grit my teeth and continue with it. However, if I were to speak as a technical expert, I would honestly say this is bad and useless.}''

\underline{R6.2 Clients want something ``cool''.}
P8 said, ``\textit{Many 3D visualizations are made because they look cool, and this has a magical appeal to clients...although they never give a definition to 'cool'.}'' P5 added, ``\textit{Me too. I have to come up with some ideas to meet the 'cool' requirement, even though they hurt scientific integrity.}'' However, P5 then supplemented, ``\textit{But I also understand my clients...from their perspective, if the visualizations I create do not go beyond their common knowledge, why would they spend money to hire me?}''  \\

\begin{wrapfigure}[3]{l}{0.035\textwidth}
\centering
\vspace*{-12pt}
\includegraphics[width=0.05\textwidth]{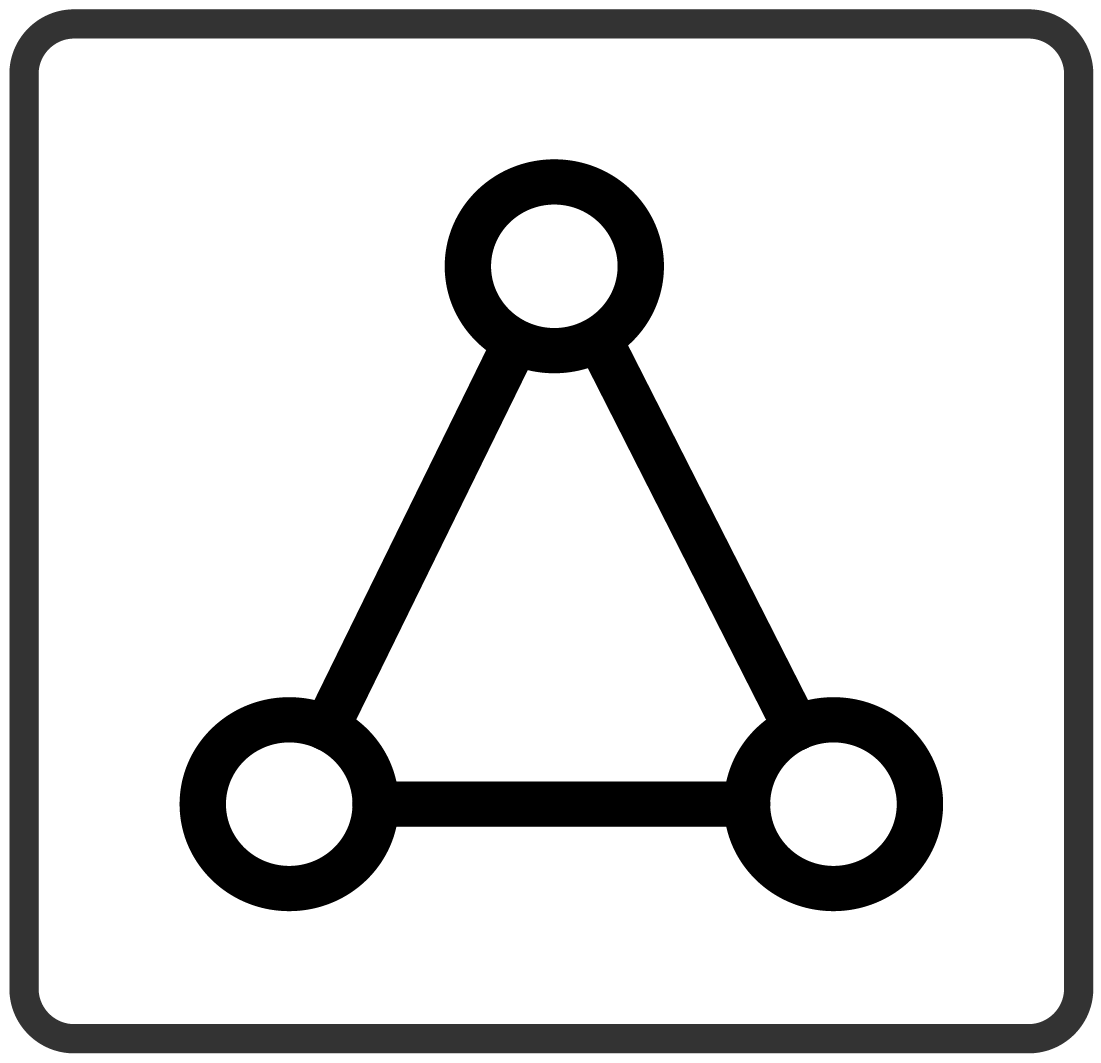}
\end{wrapfigure}
\noindent
\textbf{R7: Ineffective team collaboration.}
Two participants noted that a visualization design may be influenced by multiple roles behind the design pipeline.

\underline{R7.1 Lack of communication between multiple roles.}
P20 took her own working experience as an example and said, ``\textit{a data visualization project may involve many people working on it...insufficient communication or disagreements may arise among these people. Designers cannot make all the decisions entirely, and sometimes they may deviate from the intended goal and produce junk.}''

\underline{R7.1 Poor organization of workflow.}
As said by P16, ``\textit{In our company, designers sometimes use fictitious data to perform the visualization design, and this is because the real data has not yet been provided by a third party or the collaboration progress is not smooth. As a result, when the design is finally applied to the real dataset, the outcome is bad because the data and design are not fully compatible.}''

\subsection{Suggested Application}
The focus group studies have uncovered a variety of reasons behind visualization design flaws, ranging from intentional manipulation to unconscious or involuntary errors.
However, it should be noted that the above-mentioned points may not manifest in every circumstance. For instance, certain visualization projects may lack a specific client (R6.1) or involve less complex datasets (R4.1).
Therefore, we recommend using the above findings as heuristics, a framework or checklist that guides individuals in evaluating or reflecting on their design work~\cite{setlur2023heuristics,lan2021smile}. For example, team leaders may leverage these heuristics to better manage their visualization projects, enhancing awareness of potential risks arising from different stages and roles and taking actions to mitigate them. Additionally, designers can conduct audits of their own design work by systematically evaluating each of the identified points along with the taxonomy summarized in~\cref{sec:space}. Given that design often deals with wicked problems (\ie problems that are ill-formulated, where the information is confusing, or where there are many clients and decision-makers with conflicting values~\cite{buchanan1992wicked}), such an approach paves the way for a more structured and comprehensive identification and resolution of visualization design flaws.

\section{A research agenda for combating visualization design flaws}
\label{sec:agenda}

\begin{figure}[t]
 \centering
 \vspace{-1em}
 \includegraphics[alt={We have integrated the identified seven causes of design flaws into the information visualization reference model. From left to right: R4 happens during the data transformation stage, R5 happens during the visual mapping stage, R1 happens because of the designer's intents, R2 and R3 happens because of the designer's skills, and R6 and R7 are external factors. Below R1-R7 are nine research opportunities: O1 and O2 relate to R4, O3 and O4 relate to R5, O5 relates to R1, O6 and O7 relate to R2 and R3, and O8 and O9 relate to R6 and R7.},width=\columnwidth]{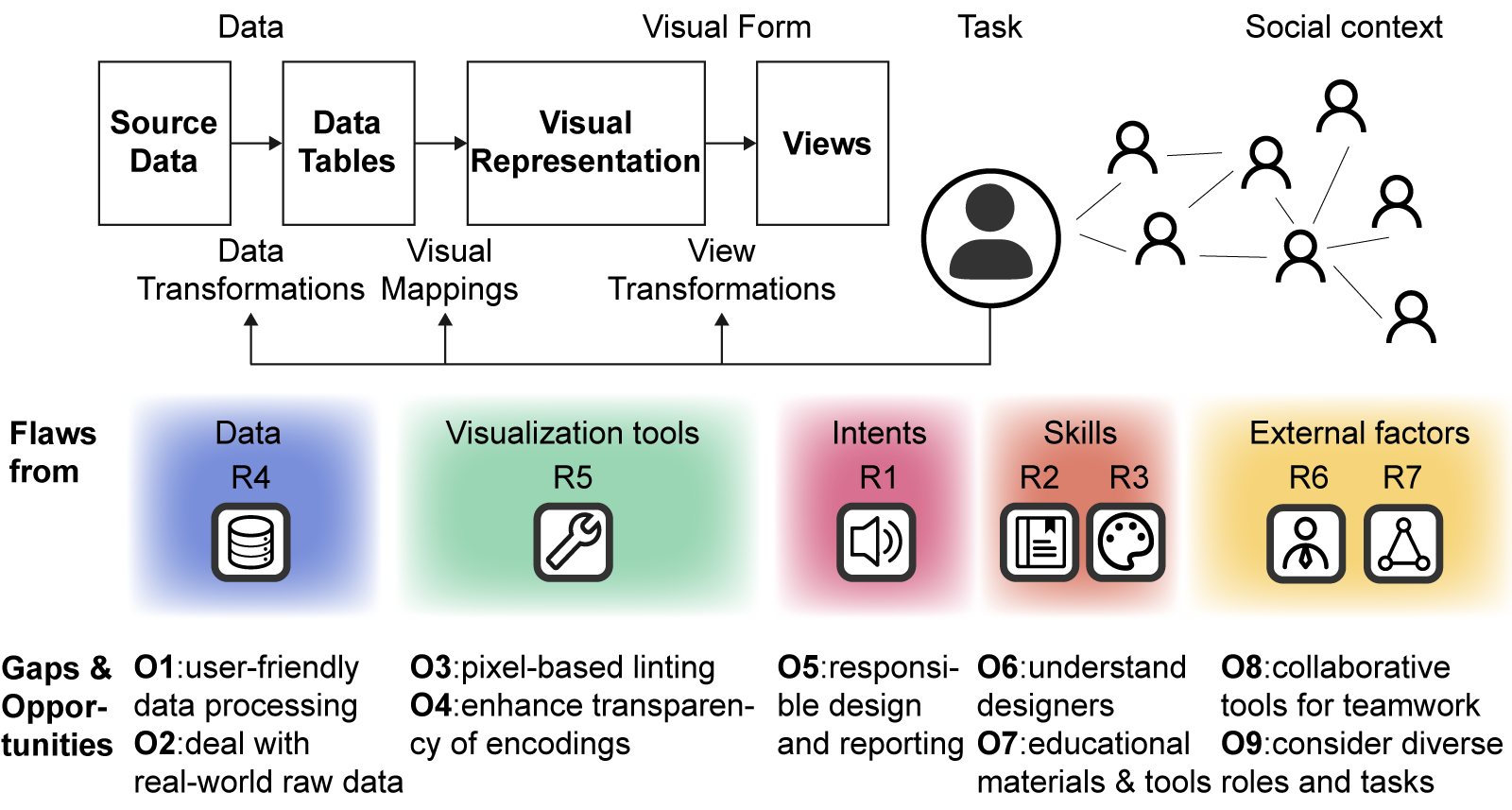}
 \caption{Locating R1-R7 identified in \cref{sec:focus} in the InfoVis reference model~\cite{card1999readings}. O1-O9 correspond to the nine research opportunities discussed in \cref{sec:agenda}.}
 \label{fig:combat}
 \vspace{-2em}
\end{figure}

Based on the findings from \cref{sec:focus}, we structured R1-R7 within the classic information visualization reference model~\cite{card1999readings}. As indicated by the color blocks in \cref{fig:combat}, visualization design flaws can stem from various stages. 
Below we discuss the tasks for combating design flaws in each stage and future research opportunities (marked from O1-O9).

\setul{0.3ex}{0.3ex}
\setulcolor{t1}
\ul{\textbf{Processing complex and untidy data.}}
Difficult-to-handle data can exacerbate the risk of design flaws. 
So far, abundant work has been conducted to assist in data wrangling and cleaning, such as automatically flagging problematic data and performing data merging~\cite{kandel2011wrangler,kandel2012profiler}. Additionally, various technologies have emerged for dimensionality reduction, enhancing data display efficiency, and avoiding visual clutter~\cite{liu2016visualizing,ellis2007taxonomy}. These technologies can effectively help users deal with dirty and high-dimensional data. However, apart from encouraging more of this type of work, we also propose that there should be more data processing technologies aimed at general users and real-world complex situations. 
For example, past research often assumes datasets are structured tables before processing, but Bartram~\etal~\cite{bartram2021untidy} showed that real-world spreadsheets frequently do not meet basic data science standards, exhibiting untidy features like merged cells and manual annotations. These non-standard situations have deep-rooted causes (\eg users need hands-on control of data). It is worth exploring how to \ul{(O1)} help users process data while accommodating their work habits and data literacy. Also, suboptimal datasets can result from data sources that are messy, scattered, and difficult to manage. Addressing this issue, Liu~\etal~\cite{liu2023governor} developed a method to assist users in searching and integrating tables from open governmental data portals, transforming unionable tables into structured sheets. Such work that focuses on \ul{(O2)} transforming real-world messy raw data into usable datasets is also highly commendable.

\setul{0.3ex}{0.3ex}
\setulcolor{t2}
\ul{\textbf{Linting and transparency in visualization tools.}}
Although the visualization community is committed to developing more technologies that detect visualization flaws, most of them are based on specialized visualization grammars (\eg Vega-Lite, matplotlib~\cite{chen2021vizlinter, mcnutt2020surfacing}). But in reality, many users do not yet possess the skills to handle such tools. A large number of in-the-wild visualizations are not code-based and are created using drawing tools such as PowerPoint and Illustrator. Therefore, there is a pressing need to \ul{(O3)} develop more linting technologies based on pixel-based inputs and make them accessible in commonly used environments. An interesting attempt in this direction comes from Fan~\etal~\cite{fan2022annotating}, who developed a method capable of identifying common mistakes in line charts directly from images and implemented it as a Google Chrome extension.
Moreover, apart from post hoc linting, \ul{(O4)} enhancing the transparency of encodings during the visualization process is also essential. For example, researchers have explored interaction techniques such as in situ annotation~\cite{hopkins2020visualint} and quasimodes~\cite{ritchie2019lie} to help users timely realize design flaws.

\setul{0.3ex}{0.3ex}
\setulcolor{t3}
\ul{\textbf{Establishing ethical guidelines.}}
Visualization design flaws can erode social trust through deliberate data manipulation. However, prior discussions on trustworthy visualization have focused more on the user side, examining how specific design choices impact trust~\cite{mayr2019trust}, yet lacking ethical guidelines for visualization creation. Addressing this gap is a complex task that requires long-term efforts from both academia and industry. But as highlighted by Correll and Heer~\cite{correll2017black}, as visualization researchers, we need to first be aware of our own responsibilities, as ``we have a great deal of power over how people ultimately make use of data''.
Therefore, we advocate for \ul{(O5)} more responsible and detailed reporting of visualization design in tools and systems. As one example, Correll~\cite{correll2019ethical} proposed a set of ethical dimensions for visualization research, noting, for instance, that ``visualization work often focuses on the positive aspects of a system (for instance, its ease of use, or the speed or accuracy with which analysts conduct their tasks), but rarely on the potential of these systems for harm or misuse.'' We believe such concrete guidelines are crucial for our community.

\setul{0.3ex}{0.3ex}
\setulcolor{t4}
\ul{\textbf{Understanding and educating designers.}}
As the barrier to creating data visualization becomes lower, the \textit{designers} we discuss today are not necessarily individuals trained in professional visualization techniques, but rather \textit{everyday designers}, namely non-experts who conduct design activities using available resources in their environment~\cite{o2016understanding}. Providing these designers with education on data visualization is a critical step in preventing design flaws. However, a significant gap we see in current research is that designers often appear in abstract forms~\cite{parsons2021understanding}. Despite numerous works claiming to serve designers, there is a lack of \ul{(O6)} in-depth examination of who these so-called designers are and how they work. For example, why do they not use professional visualization tools? Is it due to a lack of familiarity, inability to master them, or other reasons~\cite{li2024we}? What are their preferred work modes and workflows~\cite{bako2022understanding,bigelow2014reflections}? Addressing these questions calls for more research to understand designers' literacy, knowledge structures, working environments, and actual design processes. This understanding will further inspire the \ul{(O7)} development of educational materials and tools.

\setul{0.3ex}{0.3ex}
\setulcolor{t5}
\ul{\textbf{Supporting communication and collaboration with stakeholders.}}
Stakeholders can influence visualization outcomes. Some may lack expertise or fail to recognize the efforts of visualization designers. While establishing the authority of visualization professionals is important, from a more practical perspective, we believe it is worthwhile to \ul{(O8)} develop collaborative tools that support teamwork in visualization. For example, reflecting on their own visualization projects, Walny~\etal~\cite{walny2019data} identified a set of challenges that frequently happen during visualization design handoffs, such as team members' inconsistent interpretations of visualizations and data changes. They then proposed that future tools should facilitate better prototyping, testing, and communication of data-driven designs. Besides, since previous research has mainly focused on collaborative visual analytics~\cite{isenberg2011collaborative}, where users act as analysts, we suggest that future work should \ul{(O9)} expand collaborative visualization methods to accommodate diverse user roles and tasks such as monitoring, explaining, and negotiating~\cite{setlur2023heuristics}.

\section{Limitations}

This work has several limitations. Firstly, the flawed visualizations we analyzed are sourced from the WTF Visualizations gallery, where all visualizations are actively contributed by the general public. While we believe this corpus is a novel resource providing firsthand materials and opinions about visualization design flaws from users' perspectives, it is not exhaustive. For example, active contributions to this gallery are likely limited to users familiar with it or who have heard of it, potentially influenced by social connections between these users and the gallery maintainer.
Second, we analyzed these visualizations mainly through qualitative coding. This approach may inevitably introduce subjectivity from the coders. We mitigated this issue as much as possible by carefully considering the original comments of the uploaders themselves regarding the flaws they believe exist in the visualizations.
In addition, this study primarily focused on analyzing design flaws that are visible in images. Consequently, flaws occurring in earlier stages, such as data collection and cleaning, may not be easily detected through image analysis alone. This limitation may explain why perceptual flaws dominate our taxonomy and why some reasoning flaws identified in previous research (\eg errors during data curation and wrangling \cite{mcnutt2020surfacing}, cherry-picking data \cite{lisnic2023misleading}) did not emerge in this study.
Lastly, while the methodology of the focus group has helped us uncover a variety of reasons behind visualization design flaws, providing an expanded categorization of the underlying causes compared to previous research, its findings may not universally apply to every context. Additionally, we clarify that the flaw causes we identified are only post hoc rationalizations based on practitioner reviews, rather than revealing the true causes of our collected flawed samples.



\section{conclusion}

This work investigates visualization design flaws in the eyes of the general public systematically while also exploring the underlying reasons for these flaws. First, we analyzed 2227 flawed data visualizations collected from an online gallery and constructed a taxonomy containing 76 visualization design flaws. The flaws were further classified into three high-level categories (\ie misinformation, uninformativeness, unsociability) and ten subcategories. Next, we conducted five focus group studies to explore why visualization design flaws occur. Based on all the above research, we synthesized the implications of this work by proposing a research agenda for combating visualization design flaws and suggesting future research opportunities.

\acknowledgments{
This work was supported by NSFC 62402121 and Research and Innovation Projects from the School of Journalism at Fudan University. We would like to thank all the reviewers for their valuable feedback, all the participants of the focus groups who selflessly shared their experiences, and Drew Skau for hosting such a wonderful gallery and providing his unique insights for this work.}

\bibliographystyle{abbrv-doi-hyperref}

\bibliography{reference}


\end{document}